\documentclass[preprint]{ptephy_v1}

\preprintnumber{XXXX-XXXX} 
\usepackage{hyperref}
\usepackage{color}

\begin{document}

\title{Effects of Bond-Randomness and Dzyaloshinskii-Moriya Interactions on the Specific Heat at Low Temperatures of a Spherical Kagom\'{e} Cluster in \{W$_{72}$V$_{30}$\}}


\author{Mikio Motohashi}
\author[1]{Kouki Inoue}
\author[1]{Katsuhiro Morita}
\author[1]{Yoshiyuki Fukumoto}
\affil{Department of Physics, Faculty of Science and Technology, Tokyo University of Science, Noda, Chiba 278-8510, Japan \email{yfuku@rs.tus.ac.jp\;(Y. F.)}}

\author{Hiroki Nakano}
\affil{Graduate School of Science, University of Hyogo, Kamigori, Hyogo 678-1297, Japan}


\begin{abstract}%
For the spin-1/2 spherical kagom\'{e} cluster, as well as for the 2D kagom\'{e} lattice, many low-energy singlet excitations have been expected to exist in the energy region below the spin gap, 
which has been actually confirmed by Kihara {\it et al.} in their specific heat measurements 
up to 10~K in \{W$_{72}$V$_{30}$\}, for which the exchange interaction was estimated as $J=115$~K.
However, the experimental result of the specific heat can not be reproduced 
by the theoretical result in the Heisenberg model. Although the theoretical result has a peak around 2 K, the experimental one does not. To elucidate this difference, we incorporate Dzyaloshinskii-Moriya (DM) interactions and bond-randomness into the model Hamiltonian for \{W$_{72}$V$_{30}$\} and calculate density of states, entropy, and specific heat at low temperatures by using the Lanczos method. We find that DM interactions do not significantly affect the energy distribution of about ten singlet states above the ground state, which are involved in the peak structure of the specific heat around 2~K, while even 10~\% bond-randomness disperses this distribution to collapse the 2~K peak.
Kihara {\it et al.} also reported experimental specific heats under magnetic fields up to 15~T $(=0.17J)$, and found that the specific heats show almost no magnetic-field dependence,
which strongly suggests that
the bond randomness is much stronger than the magnetic fields.
For example, our calculated specific heats with 50~\% randomness reproduce the experimental ones up to about 5~K.

\end{abstract}

\subjectindex{I71, I74}

\maketitle

\section{Introduction}

\begin{figure*}[b]
\centering
\includegraphics[width=14cm]{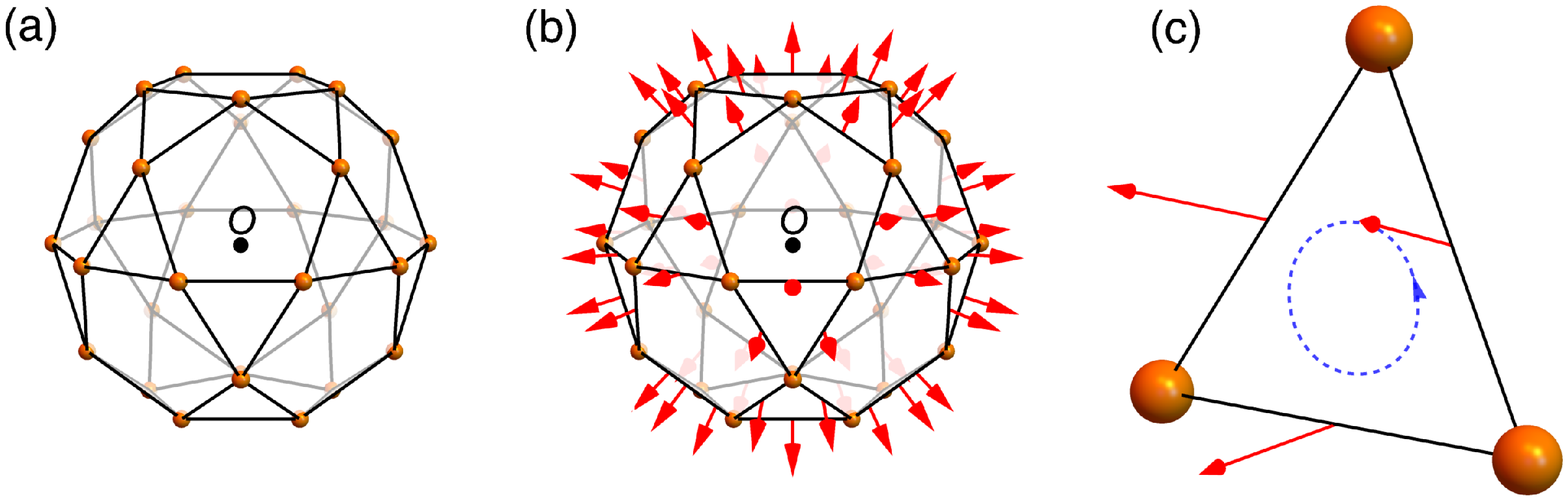}
\caption{(Color online) Schematic illustrations of (a) a spherical kagom\'{e} cluster, (b) the directions of DM interactions, and (c) the directions of bonds where $O$ is the center of the icosidodecahedron. The orange spheres, black lines, and red arrows denote $\mathrm{V^{4+}}$ ions ($S=1/2$), exchange interactions, and the directions of DM interactions respectively. The blue arrow shows the directions of bonds.}
\label{fig:W72V30}
\end{figure*}

In 1973, Anderson first proposed the resonating valence bond (RVB) states in geometrically frustrated quantum spin systems \cite{Anderson}, and then, over the past nearly half century, the Heisenberg antiferromagnets with spin-1/2 in the kagom\'{e} lattice, which is a two-dimensional network of corner-sharing triangles, have attracted a lot of attention because of the strong frustrated effect \cite{{Waldtmann1998},{Mambrini2000},{Matan2006},{Yildirim2006},{HN_TSakai_ramp_2010},{Matan2010},{Nakano2011_2},{Yan2011},{Lu2011},{Depenbrock2012},{Nishimoto2013},{Ono2014},{Fu2015},{Shimokawa2015},{HNakano_kgm45},{Saito2021}}. We here focus on spin-1/2 spherical kagom\'{e} clusters \cite{{Hasegawa2004},{Rousochatzakis},{Kunisada2008},{Schnack1},{Kunisada},{Kunisada2015},{Yokoyama},{Inoue},{Inoue_cTPQ_hinetu}}, or spin-1/2 icosidodecahedra, which correspond to a zero-dimensional counterpart of the kagom\'{e} lattice. An icosidodecahedron is composed of corner-sharing triangles [see Fig.~\ref{fig:W72V30}(a)] and realized in \{W$_{72}$V$_{30}$\} \cite{{Todea},{Schnack2},{Nojiri},{Kihara}}. The main part of the Hamiltonian for \{W$_{72}$V$_{30}$\} is presented as
\begin{align}
\label{Heisenberg}
\mathcal{H}=J\sum_{\langle i,j\rangle}\boldsymbol{S_\textit{i}}\cdot \boldsymbol{S_\textit{j}},
\end{align}
where $\langle i,j\rangle$ and $J$ denote nearest neighbors and exchange couplings respectively, and $\boldsymbol{S_{\textit{i}}}=(S_{i}^{x},S_{i}^{y},S_{i}^{z})$ is a spin-1/2 operator at site $i$. 

Todea {\it et al.} synthesized \{W$_{72}$V$_{30}$\} and measured its magnetic susceptibility $\chi$ in 2009 \cite{Todea}. Schnack, Kunisada, and Fukumoto calculated the magnetic susceptibility of \{W$_{72}$V$_{30}$\} with $J=115$ K \cite{{Schnack1},{Kunisada}} and reproduced the experimental result, but the measurement by Todea {\it et al.} was not carried out at very low temperatures. Thus, it was unclear whether the agreement was consistent even at very low temperatures below $0.1J \simeq 10$ K. Also, Schnack, Kunisada, and Fukumoto calculated the magnetization process and the specific heat \cite{{Schnack1},{Kunisada}}, which were expected to be observed experimentally in \{W$_{72}$V$_{30}$\}.

However, contrary to the theoretical prediction, later experiments show that although the theoretical magnetization curve has staircase structure the experimental one increases linearly with magnetic field and that theoretical susceptibility vanishes faster than measured susceptibility as the temperature drops to low \cite{{Schnack2},{Nojiri},{Kihara}}. Schnack {\it et al.} suggested that the discrepancies were attributed to the distribution of nearest-neighbor exchange couplings, which was called bond-randomness, where the width of the variance in exchange interactions was estimated to be 30 \% of the average value $J$ \cite{Schnack2}.
Since the triplet excitation gap of the Heisenberg model in Eq.~(\ref{Heisenberg}) is $\Delta_{\mathrm{t}}=0.218J$ \cite{Rousochatzakis}, the collapse of the staircase structure in the magnetization curve indicates that the 30 \% bond-randomness can disperse the distribution of states over an energy range $\sim 0.2J$.
On the other hand, it was confirmed that the incorporation of Dzyaloshinskii-Moriya (DM) interactions 
with the strength of $\sim 0.1J$
also leads to the collapse of the staircase structure \cite{{Yokoyama},{Inoue}}. Since the addition of either bond-randomness or DM interactions yields similar results in the susceptibility and magnetization process at low temperatures, which are both measures of magnetic states, from these results alone, it is difficult to distinguish the effects of the two perturbations. Thus, there was interest in the specific heat measurement where nonmagnetic singlets could be observed.

Subsequently, in 2019, Kihara {\it et al.} experimentally measured the specific heat of \{W$_{72}$V$_{30}$\} below 10~K $(=0.087J)$ and revealed the existence of many low-energy nonmagnetic singlet states below the first excited triplet state \cite{Kihara}. As expected from the above mentioned experiments, which suggests the existence of the DM interaction or bond-randomness, their experimental result on the specific heat also extremely differs from theoretical results for the Heisenberg model \cite{{Schnack1},{Kunisada}}. Although a clear peak around 2 K appears and the specific heat depends on the applied field in calculated results, the 2 K peak vanishes and the magnetic field dependence hardly exists in the experimental result, where the maximum applied field is $H=15$~T ($=0.17J$).
 In this work, to clarify the cause of these discrepancies, we add bond-randomness or DM interactions to Eq.~(\ref{Heisenberg}) and calculate the specific heat by using Lanczos method. Unlike the susceptibility or magnetization process, 
specific heat is determined only by the energy-eigenvalue distribution, i.e. density of states (DOS),
thus it might be possible to determine the impacts of each perturbation from the calculation of the specific heat.
It should be stressed that the major difference between bond randomness and DM interaction is that the former disperses the DOS, while the latter does not.

This paper is organized as follows. 
In Sec.~2, we introduce our model Hamiltonian and describe the detailed conditions of our calculation.
Our calculation results are shown in Sec.~3. 
In Sec.~3.1, it is shown that the 2~K peak does not disappear even though DM interactions are added.
In Sec.~3.2, we show even $10$~\% bond-randomness leads to collapse of the 2~K peak, but we find the $10$~\% bond randomness does not weaken the magnetic field dependence enough and does not reproduce the experiment well.
Section~3.3 is devoted to the study of larger bond-randomness, where we actually show bond randomness much stronger than the applied magnetic field makes the specific heat independent from the applied field.
A summary of our findings is shown in Sec.~4.

\section{Model Hamiltonian and calculation method}

We write the DM interaction as
\begin{align}
\label{DM}
\mathcal{H}_{\mathrm{DM}}=\sum_{\langle i,j\rangle}\boldsymbol{D_{\textit{i,j}}}\cdot(\boldsymbol{S_\textit{i}}\times \boldsymbol{S_\textit{j}}),
\end{align}
which originates from the spin-orbit coupling. Since it has the form of a cross product, we need to define the direction of a bond $\langle i,j \rangle$. We order the three sites in each triangle counterclockwise from the outside of an icosidodecahedron [see Fig.~\ref{fig:W72V30}(c)] and regard site $i$ ($j$) as the prior (subsequent) site. Since there exists a mirror plane perpendicular to the bond direction for a bond $\langle i,j \rangle$, $\boldsymbol{D_{\textit{i,j}}}$ is lying in the mirror plane. We introduce the two unit vectors lying in the mirror plane, 
\begin{align}
\boldsymbol{e_{\textrm{r}}}^{(i,j)}=\frac{\boldsymbol{r_\textit{i}}+\boldsymbol{r_\textit{j}}}{|\boldsymbol{r_\textit{i}}+\boldsymbol{r_\textit{j}}|}\ \ \mathrm{and}\ \ \boldsymbol{e_{\textrm{p}}}^{(i,j)}=\frac{\boldsymbol{r_\textit{i}}\times \boldsymbol{r_\textit{j}}}{|\boldsymbol{r_\textit{i}}\times \boldsymbol{r_\textit{j}}|},
\end{align}
where $\boldsymbol{r_\textit{i}}\ (\boldsymbol{r_\textit{j}})$ denotes the position vector for the site $i$ ($j$) and the center of the icosidodecahedron is chosen as the origin $O$. Then, on the basis of a symmetry consideration \cite{Yokoyama}, we can obtain the following expression
\begin{align}
\mathcal{H}_{\mathrm{DM}}=D\sum_{\langle i,j\rangle}(\cos{\theta}\,\boldsymbol{e_{\textrm{p}}}^{(i,j)}+\sin{\theta}\,\boldsymbol{e_{\textrm{r}}}^{(i,j)})\cdot(\boldsymbol{S_\textit{i}}\times \boldsymbol{S_\textit{j}}),
\end{align}
where $\theta$ is the angle defining its direction and $D=|\boldsymbol{D_{\textit{i,j}}}|$. Fukumoto {\it et al.} found that the direction of $\boldsymbol{D_{\textit{i,j}}}$ parallel to the radial direction efficiently cancels out the staircase behavior of the magnetization curve in the low magnetic field region and that the tendency begins to appear with the addition of at least 10 \% of DM interactions \cite{Yokoyama}. From this result, we adopt $\theta=
0.5\pi$ and $D=0.1J$ in our calculation. The directions of DM interactions are schematically shown in Fig.~\ref{fig:W72V30}(b).

The Hamiltonian describing the bond-randomness is
\begin{align}
   \mathcal{H}_{\mathrm{Random}}=&\sum_{\langle i,j\rangle}\alpha_{i,j}\ \boldsymbol{S_\textit{i}}\cdot \boldsymbol{S_\textit{j}},
\end{align}
where $\alpha_{i,j}$ is a uniform random value between $-\Delta J$ and $\Delta J$. 
We set the size of bond-randomness as 10 \%, that is, $\Delta J=0.1J$, in Sec~3.2.
The first singlet excitation gap of the Heisenberg model in Eq.~(\ref{Heisenberg}) was estimated to be $\Delta_{\mathrm{s}}=0.048J\sim \frac{1}{4}\Delta_{\mathrm{t}}$ \cite{Kunisada}, and thus, the 10 \% bond-randomness is enough to disperse the distribution of states over an energy range $\sim \Delta_{\mathrm{s}}$, which is plausible to give an impact on the specific heat at very low temperatures.
In Sec~3.3, we also present calculated results for $\Delta J=0.3J$ and $0.5J$, which are several times as large as
applied magnetic fields in the specific-heat measurement.

Now, with these perturbations added, we write our total Hamiltonian as
\begin{align}
\nonumber
\label{DM_total}
\mathcal{H}_{\textrm{DM total}}=&~J\sum_{\langle i,j\rangle}\ \boldsymbol{S_\textit{i}}\cdot \boldsymbol{S_\textit{j}} 
+D\sum_{\langle i,j\rangle}(\cos{\theta}\ \boldsymbol{e_{\textrm{p}}}^{(i,j)}+\sin{\theta}\ \boldsymbol{e_{\textrm{r}}}^{(i,j)})\cdot(\boldsymbol{S_\textit{i}}\times \boldsymbol{S_\textit{j}}) \\
&-g\mu_{\mathrm{B}}\boldsymbol{H}\cdot \sum_{i}\boldsymbol{S_\textit{i}},
\end{align}
or
\begin{align}
\label{rnd_total}
\mathcal{H}_{\textrm{Random total}}=&\sum_{\langle i,j\rangle}(J+\alpha_{i,j})\ \boldsymbol{S_\textit{i}}\cdot \boldsymbol{S_\textit{j}}-g\mu_{\mathrm{B}}\boldsymbol{H}\cdot \sum_{i}\boldsymbol{S_\textit{i}},
\end{align}
where $-\Delta J\leq \alpha_{i,j} \leq\Delta J$ ($\Delta J=0.1J$, $0.3J$ or $0.5J$), $D=0.1J$, $\theta=0.5\pi$, $g=1.95$, and $J=115$~K, and $\boldsymbol{H}$ denotes the magnetic field. 
We use $|\boldsymbol{H}|=$ 0~T and 10~T $(=0.11J)$ when quantifying the magnetic field-induced change in specific heat.
The orientations of magnetic clusters in polycrystalline samples of \{W$_{72}$V$_{30}$\} used in the magnetization measurements \cite{{Schnack2},{Nojiri},{Kihara}}, where discrete and well-separated magnetic spherical kagom\'{e} clusters are embedded in a nonmagnetic environment, are distributed randomly and are expected to be unaffected by the magnetic field direction. Therefore, we set the direction of the magnetic field $\boldsymbol{H}$ in Eq.~(\ref{DM_total}) to be uniformly random. Also, in the synthesizing process of \{W$_{72}$V$_{30}$\} samples, each molecule embedded in a nonmagnetic environment can be somewhat distorted from the idealized perfect icosidodecahedron, and thus, the strength of exchange interactions of each bond can be fluctuated. This is thought to be one of origins of bond-randomness in \{W$_{72}$V$_{30}$\}.

We use Otsuka's calculation method \cite{Otsuka}, in which DOS is first calculated by using the Lanczos method in conjunction with a sampling technique and then thermodynamic quantities are obtained via the DOS. Since we treat a large scale matrix whose dimension of the Hilbert space is $2^{30}$, our numerical calculations were performed on the ISSP system B (ohtaka) at the Supercomputer Center, Institute for Solid State Physics, University of Tokyo, using OpenMP parallelization with up to 128 cores.

We calculate the specific heat by Otsuka's method \cite{Otsuka}, in which the random sampling basis are used. We denote the number of random sampling basis by $N_s$. A physical quantity, such as DOS, specific heat, or entropy, is obtained from this $N_s$ sampling basis. In order to evaluate the ambiguity stemming from the sampling procedure, we calculate the standard error using Seki and Yunoki's method \cite{SekiYunoki}. We conclude $N_s=20$ is sufficient for our present purpose, examining the 2 K peak and magnetic field dependence of the specific heat. We also confirmed that the drastic change of the behavior of the specific heat does not happen with $N_s$. Therefore, in the following sections, we adopt $N_s=20$. 
For the Hamiltonian in Eq.~(\ref{DM_total}) with $H=|\boldsymbol{H}|>0$, we repeated the specific heat calculation 5 times, in each of which the direction of the magnetic field is chosen independently, and averaged over 5 data obtained here. In this calculation it was observed that the distribution arising from the magnetic-field direction is much smaller than that arising from the random sampling of basis vectors.

We note that the specific heat of the Heisenberg model for \{W$_{72}$V$_{30}$\} using the Lanczos method in the present study is good agreement with that obtained using canonical Thermal Pure Quantum method in a previous study \cite{Inoue_cTPQ_hinetu}. The reason why we used Otsuka's method in the present study is to obtain DOS, which helps us to understand the behavior of the specific heat.

\section{Results and discussions}

As mentioned in the introduction, the key differences of the specific heat between the experimental result and the theoretical result in \{W$_{72}$V$_{30}$\} are the existence of the peak around 2 K and the magnetic field dependence. Thus, we discuss our calculation results from these two points mainly.

\subsection{DM result (the case of Eq.~(\ref{DM_total}))}

\begin{figure}[t]
\centering
\includegraphics[width=13.5cm]{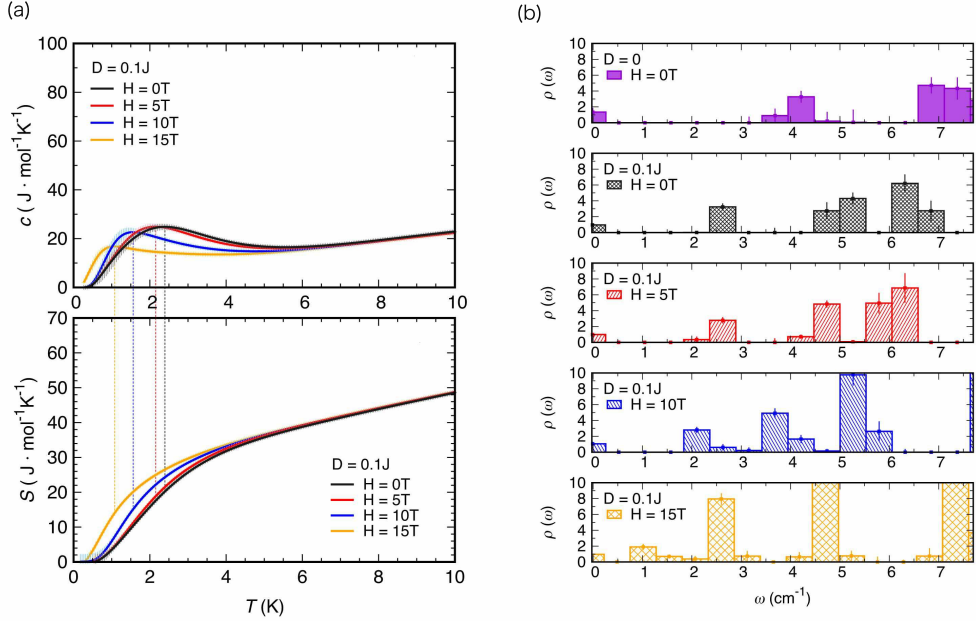}
\caption{(Color online) 
(a) Temperature dependence of specific heat and entropy for $D=0.1J$ at $H=$ 0, 5, 10, 15 T $(=0.17J)$. 
Dashed lines in the figure represent the peak temperature of the specific heat and the entropy corresponding to that temperature. 
(b) Calculated DOS around the ground state at $H=$ 0, 5, 10, 15 T  for $D=0.1J$ and at $H=$ 0 T for $D=0$.
}
\label{newfig2}
\end{figure}

We first study the effect of DM interactions by using the Hamiltonian in Eq.~(\ref{DM_total}).
In Fig.~\ref{newfig2}(a), we show our calculated result of the specific heat and entropy at $H=$ 0, 5, 10, 15~T $(=0.17J)$ below the temperature of 10~K $(=0.087J)$.
The upper panel shows that the 2 K peak still remains even though DM interactions are incorporated, 
and the lower panel shows that the values of each entropy corresponding to the peak temperature of the specific heat at $H=$ 0, 5, 10, 15~T are $S=$ 20, 19, 15, and 14 $\textrm{J}\cdot \textrm{mol}^{-1}\textrm{K}^{-1}$ respectively, which are comparable to 11, 10, 6, and 5 states from the Boltzmann principle. Thus, we can realize that about ten or a little less states counting from the ground state determine the structure up to the 2 K peak of the specific heat. The presence of the 2 K peak implies that the addition of DM interactions does not significantly affect the distribution of ten or a little less states from the ground state.

Also, Fig.~\ref{newfig2}(b) shows the DOS result, where the unit of energy is cm$^{-1}$. 
(In this unit, Eq.~(\ref{Heisenberg}) has the first triplet gap $\Delta_{\mathrm{t}}=0.218J=17.4$ cm$^{-1}$ and the first singlet gap $\Delta_{\mathrm{s}}=0.048J=3.7$ cm$^{-1}$.) 
The width of the first excitation gap in Fig.~\ref{newfig2}(b) seems to be almost unchanged when the applied field is low while it gets narrower at $H=$ 15 T.
From Fig.~\ref{newfig2}(b), it is seen that the number of states at $H=$ 0 T exceeds 11 at $\omega=5.1\ \textrm{cm}^{-1}$, and at $H=$ 5 T, the number of states exceeds 10 at $\omega=4.8\ \textrm{cm}^{-1}$. The 2 K peak position at $H=$ 0, 5 T is almost the same because the distribution of about 10 states counting from the ground state is roughly the same structure. On the other hand, the number of states exceeds 6 at $\omega=3.6\ \textrm{cm}^{-1}$ when $H=$ 10 T and 5 at $\omega=2.4\ \textrm{cm}^{-1}$ when $H=$ 15 T. Since for these two the number of states reach the total number of states determining the peak structure of the specific heat faster than those at $H=$ 0 T and 5 T, each peak position moves toward the left depending on how fast they reach the total number of states that lead to the peak structure.

\begin{figure}[t]
\centering
\includegraphics[width=13cm]{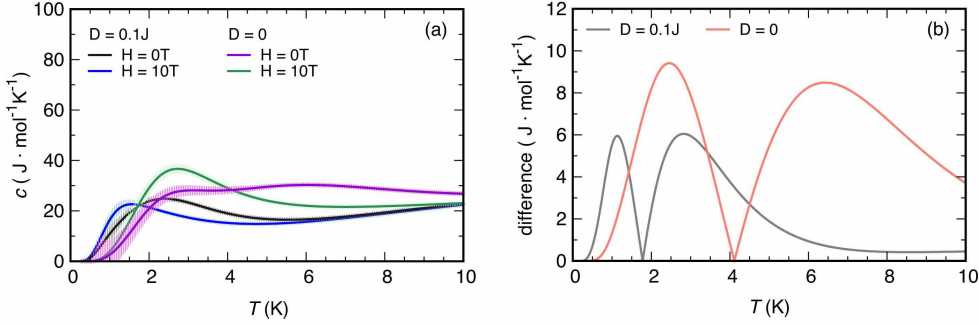}
\caption{(Color online) Temperature dependence of (a) specific heat at $H=$ 0, 10 T for $D=0$, $0.1J$, 
and (b) difference of specific heat curves at $H=$ 0~T and 10~T, where error bars are omitted.
In (b), average values of the difference for $D=0.1J$ and $D=0$ are, respectively, 
$\langle|\Delta c|\rangle=$
$2.2\ \textrm{J}\cdot \textrm{mol}^{-1}\textrm{K}^{-1}$ and $5.6\ \textrm{J}\cdot \textrm{mol}^{-1}\textrm{K}^{-1}$.
}
\label{fig:heisen+dm_hinetu_sa}
\end{figure}

Figure~\ref{fig:heisen+dm_hinetu_sa} presents 
the calcurated results of the specific heat $c$
at $H=$ 0, 10 T for $D=0$ and $D=0.1J$. From Fig.~\ref{fig:heisen+dm_hinetu_sa}(b), we find that the maximum difference between $H=$ 0 T and 10 T is $9.4\ \textrm{J}\cdot \textrm{mol}^{-1}\textrm{K}^{-1}$ for $D=0$ while it is $6.0\ \textrm{J}\cdot \textrm{mol}^{-1}\textrm{K}^{-1}$ for $D=0.1J$ and that the average of the difference,
which is defined by
\begin{equation}
   \langle|\Delta c|\rangle\equiv\frac{1}{T_{\rm{max}}}\int_{T<T_{\rm{max}}}|c_{H=10{\rm T}}(T)-c_{H=0{\rm T}}(T)|dT
\end{equation}
with $T_{\rm{max}}=10{\rm K}$,
is $\langle|\Delta c|\rangle=5.6\ \textrm{J}\cdot \textrm{mol}^{-1}\textrm{K}^{-1}$ for $D=0$ while $\langle|\Delta c|\rangle=2.2\ \textrm{J}\cdot \textrm{mol}^{-1}\textrm{K}^{-1}$ for $D=0.1J$. 
Hence, it seems that the addition of DM interactions slightly weakens the magnetic field dependence. This suggests that the DM interaction in \{W$_{72}$V$_{30}$\} suppresses the dependence of energy eigenvalues on the magnetic field.

Here, it is instructive to look into a dimer model with the DM interaction,
\begin{align}
\mathcal{H}_{12}=J\boldsymbol{S_{\textrm{1}}}\cdot \boldsymbol{S_{\textrm{2}}}+\boldsymbol{D}\cdot (\boldsymbol{S_{\textrm{1}}}\times \boldsymbol{S_{\textrm{2}}})
\end{align}
with a DM vector $\boldsymbol{D}=(D_{x},D_{y},D_{z})$. The DM interaction causes the magnetic component to mix the singlet state, and then, it seems that this mixing makes thermodynamic quantities be more sensitive to the magnetic field. However, this is not always the case, and the dimer system is a good example for it.
In the case of $D=0$, the energy eigenvalues of $\mathcal{H}_{12}$, as commonly known, are given as $\frac{J}{4}, \frac{J}{4}, \frac{J}{4}, \frac{-3J}{4}$, with the first three eigenvalues corresponding to the triplet and the last one corresponding to the singlet. Then, when $D\neq 0$, the eigenvalues of $\mathcal{H}_{12}$ are given as $\frac{J}{4}, \frac{J}{4}, \frac{-J}{4}+\frac{1}{2}\sqrt{J^2+D^2}, \frac{-J}{4}-\frac{1}{2}\sqrt{J^2+D^2}$, which are independent from the direction of $\boldsymbol{D}$ because of the rotational symmetry of the exchange term. Now, we turn to the case where the magnetic field is added as a perturbation. The introduction of the magnetic field makes the energy eigenvalues depend on the direction of $\boldsymbol{D}$. Defining the magnetic field direction of the z-axis, we obtain the Hamiltonian
\begin{align}
\mathcal{H}'_{12}=\mathcal{H}_{12}-H(S_{1}^{z}+S_{2}^{z})\equiv \mathcal{H}_{12}-HV,
\end{align}
where $H$ is the strength of the magnetic field and we regard $V=S_{1}^{z}+S_{2}^{z}$ as a perturbation operator. In the case of $D=0$, the energy eigenvalues are given as $\frac{J}{4}\pm H, \frac{J}{4}, \frac{-3J}{4}$. It is seen that the degeneracy of the three triplet is resolved by the magnetic field perturbation, and here, the coefficient of the Zeeman splitting is 2.
In the case of $D\neq 0$, first, when we add the magnetic field perturbation to the non-degenerated eigenvalues $\frac{-J}{4}\pm \frac{1}{2}\sqrt{J^2+D^2}$, the first-order perturbation energy become 0, and hence, the presence or absence of the DM interaction does not contribute to the magnetic field dependence of these two eigenvalues. Next, when we add a perturbation of the magnetic field to the degenerated eigenvalues $\frac{J}{4}$, we get $\frac{J}{4}\pm \frac{D_z}{D}H$, corresponding to $\frac{J}{4}\pm H$ for $D=0$ and its coefficient of the Zeeman splitting is given as $2\times\frac{|D_z|}{D}$, which becomes smaller than 2. In this case, when we take a directional average over $D_z$, the coefficient of the Zeeman splitting becomes $\frac{4}{\pi}\sim1.27<2$. Therefore, in this dimer model, the incorporation of the DM interaction causes the dependence of energy eigenvalues on the magnetic field to slow down. The reason for it is that the slope of the eigenvalue curve against the magnetic field becomes slower due to the level repulsion caused by the breaking of the conservation of total $S^z$ by the DM interaction. It might be expected that this mechanism based on the level repulsion is present not only in the dimer system but also in general systems.

\subsection{Small bond-randomness result (the case of Eq.~(\ref{rnd_total}) with $\Delta J=0.1J$)}

Next, we study the effect of the bond-randomness by using Eq.~(\ref{rnd_total}). There are 60 $J$-bonds in an icosidodecahedron, thus 60 $\alpha_{i,j}$'s make a sample of distorted Hamiltonian with the bond-randomness. We assign uniform random numbers $\alpha_{i,j}\ (-0.1J\leq \alpha_{i,j} \leq 0.1J)$ to make 5 samples of the distorted Hamiltonians.
%
%
Thermodynamic quantities calculated from a sampled Hamiltonian tend to depend on the values of $\alpha_{i,j}$'s. Since we use the exact diagonalization method in our calculation, the number of samples we can prepare is practically very limited. Under such circumstances, we intend to extract properties common to all samples, which, we expect, leads us to qualitative understanding of the effect of bond-randomness for \{W$_{72}$V$_{30}$\}.

\begin{figure}[t]
\centering
\includegraphics[width=12cm]{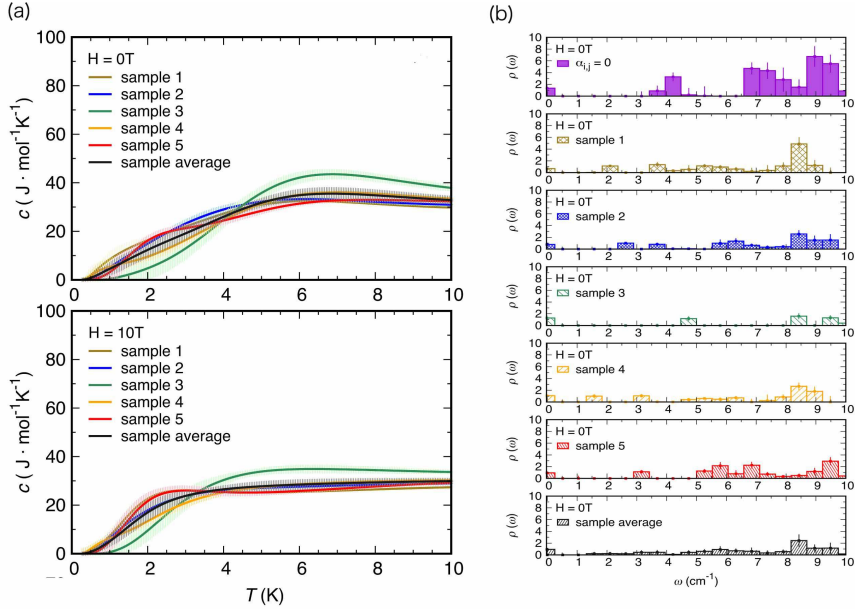}
\caption{(Color online) (a) Specific heat for $-0.1J\leq \alpha_{i,j} \leq 0.1J$ at $H=$ 0 T and 10 T.
Each sample number in the figure denotes a distorted Hamiltonian determined from 60 combinations of uniform random numbers $\alpha_{i,j}$.
(b) The DOS at $H=$ 0 T. The data of ``$\alpha_{i,j}=0$" corresponds to that of ``$D=0$, $H=$ 0 T" in Fig.~\ref{newfig2}(b).
}
\label{fig:newfig4}
\end{figure}

Figure~\ref{fig:newfig4}(a) shows the specific heat at $H=$ 0 T and $H=$ 10 T. 
From Fig.~\ref{fig:newfig4}(a) we find that the 2 K peak of the specific heat vanishes in contrast to the DM result.

\begin{figure}[t]
\centering
\includegraphics[width=14cm]{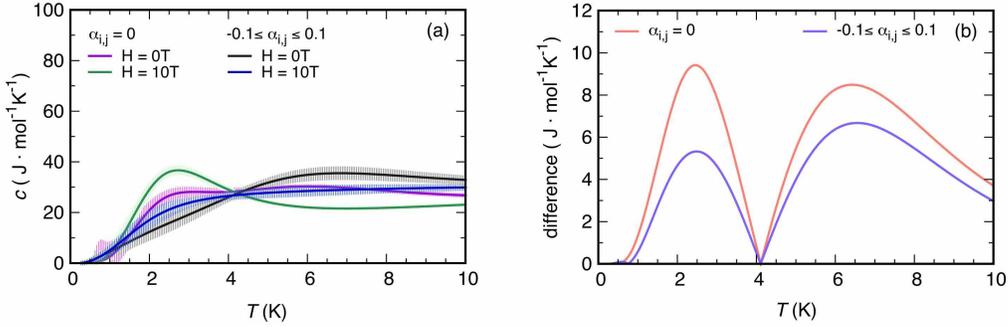}
\caption{(Color online) The temperature dependence of (a) specific heat at $H=$ 0, 10T $(=0.11J)$ for $\alpha_{i,j}=0$ and $-0.1J\leq \alpha_{i,j} \leq 0.1J$,
which is the average of 5 samples in Fig.~\ref{fig:newfig4}(a),
and  (b) the difference of specific heat between $H=$ 0 T result and $H=$ 10 T result of (a), where error bars are omitted.
In (b), average values of the difference,
$\langle |\Delta c|\rangle$, 
are $4.0\ \textrm{J}\cdot \textrm{mol}^{-1}\textrm{K}^{-1}$ for 10\% randomness and $5.6\ \textrm{J}\cdot \textrm{mol}^{-1}\textrm{K}^{-1}$ for 0\% randomness, respectively.
}
\label{fig:random_hinetu_sa}
\end{figure}

Remembering the discussion given in Sec. 3.1, we expect ten or a little less states involved in the peak structure of the specific heat to be dispersed by the bond-randomness effect. The DOS at $H=$ 0 T are presented in Fig.~\ref{fig:newfig4}(b). Figure~\ref{fig:newfig4}(b) shows that the DOS distributions of each sample are quite varied, and we can see the flatted DOS distribution in the average of 5 samples. Thus, it is concluded that the DOS distribution is dispersed by the addition of bond-randomness, which results in the disappearance of the 2 K peak.

In Fig.~\ref{fig:random_hinetu_sa}, we show the specific heat results at $H=$ 0, 10 T for $\alpha_{i,j}=0$ and $-0.1J\leq \alpha_{i,j} \leq 0.1J$. From Fig.~\ref{fig:random_hinetu_sa}(b), it is found that the maximum difference between $H=$ 0 T and 10 T is $6.7\ \textrm{J}\cdot \textrm{mol}^{-1}\textrm{K}^{-1}$ and 
the average of the difference is 
$\langle |\Delta c|\rangle=$
$4.0\ \textrm{J}\cdot \textrm{mol}^{-1}\textrm{K}^{-1}$ for $-0.1J\leq \alpha_{i,j} \leq 0.1J$. This result suggests that, compared to the DM result
of $\langle |\Delta c|\rangle=2.2\ \textrm{J}\cdot \textrm{mol}^{-1}\textrm{K}^{-1}$ for $D=0.1J$,
the magnetic field dependence of the specific heat is not weakened so much when the 10\% bond-randomness is added. Also, from Fig.~\ref{fig:random_hinetu_sa}(a), it is seen that the shoulder of the specific heat moves from $T\simeq 6\ \textrm{K}$ at $H=$ 0 T toward $T\simeq 3\ \textrm{K}$ at $H=$ 10 T when bond-randomness is incorporated. The curves in Fig.~\ref{fig:random_hinetu_sa}(a) for $\alpha_{i,j}=0$ show the shoulder at 6 K shifts to the left by applying the field, which is the same response in Fig.~\ref{fig:random_hinetu_sa}(a) for $-0.1J\leq \alpha_{i,j} \leq 0.1J$. Hence, 
small bond-randomness does not have much of an effect on the magnetic field dependence. 

\begin{figure}[h]
\centering
\includegraphics[width=15.5cm]{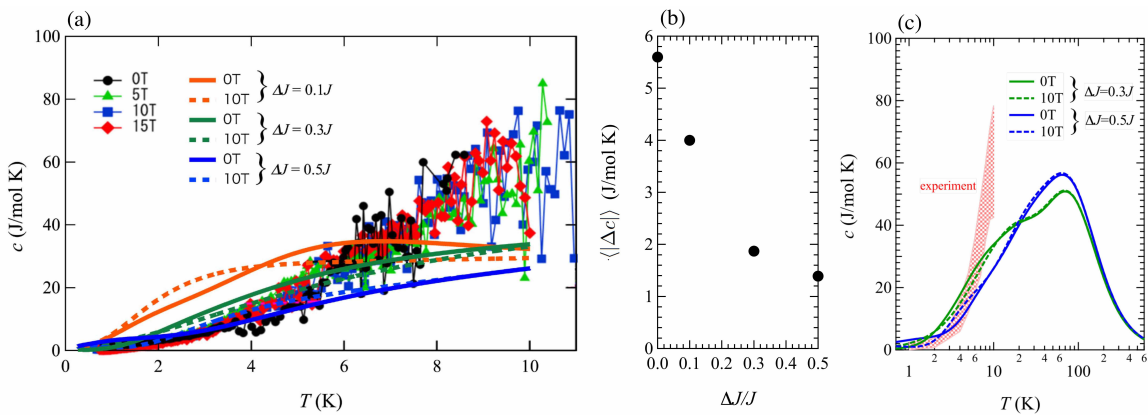}
\caption{
(Color online) (a) Temperature dependence of calculated specific heat for the 10\%, 30\% and 50\% bond-randomnesses.
Markers show experimental results of \{W$_{72}$V$_{30}$\} reported by Kihara {\it et al.} \cite{Kihara}.
(b) Average difference 
$\langle|\Delta c|\rangle$ 
as a function of $\Delta J/J$.
(c) Specific heat curves for the whole temperature.
}
\label{fig:hikaku}
\end{figure}

In Fig.~\ref{fig:hikaku}(a), we compare our calculated specific heat for the 10\% bond-randomness, shown by the orange line, with the experimental data.
Although the 2 K peak is wiped out by the 10\% randomness, the dependence on magnetic field is pronounced and the initial slopes are much larger than experiment.
Thus, in the next subsection, we study how larger bond-randomness affects the temperature dependence of specific heat.

\subsection{Large bond-randomness result (the case of Eq.~(\ref{rnd_total}) with $\Delta J=0.3J$ and $0.5J$)}

We here present our calculated results for laregr bond-randomnesses.
In this context, Watanabe {\it et al.} studied the random Heisenberg antiferromagnet on the triangular lattice to understand quantum spin liquid behaviour observed in organic salts, 
including temperature-linear specific heat which is insensitive to magnetic fields \cite{Kawamura}. 
To be precise, they presented calculated results of specific heat for 100\% bond randomness, i.e. $\Delta J=J$, on a triangle cluster with the size of $N=18$,
and showed that magnetic fields of $H=0,\;0.05J,\;0.1J$ do not change the temperature dependence of the specific heat
(see Fig.~4(a) in \cite{Kawamura}).
Thus, it is interesting to study our spherical Kagom\'{e} cluster in the case of $\Delta J \gg H$.

Since the applied magnetic field in the experiment is about $H=10{\rm T}=0.11J$, we consider $\Delta J=0.3J$ and $0.5J $ which satisfy $\Delta J \gg H $.
However, we find, as the randomness increased, the variation in the specific heat of the zero magnetic field increased depending on the sample, 
so the average of the zero magnetic field case is taken for 10 samples.
We plot the average difference 
$\langle|\Delta c|\rangle$ 
as a function of $\Delta J/J$ in Fig.~6(b),
which shows that the magnetic field dependence is suppressed as the randomness increases.

In Fig.~6(a), we show a comparison between the experimental data of specific heat and the calculation results at 30\% and 50\% randomnesses.
The initial slope becomes smaller as the randomness increases, and the specific heat of 50\% randomness is consistent with the experiment at 5K or less.
Specific heat curves for the whole temperature are shown in Fig.~6(c), in which the red-hatched region represents the experimental specific heat below 10 K.
Figure 6(c) indicates that the difference in the experimental and calculated results around $T=10 {\rm K}$ is rather significant, 
suggesting the contribution of degrees of freedom other than the spin degree of freedom.
The main energy scale of the present system is $J=115 {\rm K}$, and the Heisenberg model with it is known to reproduce, in the range of $10 {\rm K} - 300 {\rm K}$, the experimental result of spin susceptibility,
which, in general, can be observed over a wide temperature range without being masked by other degrees of freedom.
On the other hand, the lattice specific-heat gives a $T^3$ term to the total specific-heat, and thus, the spin specific-heat is easily masked by the lattice specific-heat when the temperature rises.
It is natural to interpret the difference between the calculated and experimental spin specific-heats around 10 K in Fig.~6(c) as being due to an ambiguity in the subtraction of the lattice specific-heat from the total experimental specific-heat. 

\section{Summary and future problems}

We have investigated the impacts of DM interactions and bond-randomness on the specific heat of \{W$_{72}$V$_{30}$\} by using the Lanczos method. 
We have found that the peak of the specific heat around 2 K still remains even though DM interactions are incorporated because DM interactions do not have much effect on the DOS distribution leading to the peak structure. 
Also, DM interactions tend to reduce the magnetic field dependence. 
On the other hand, the 10\% distribution of nearest-neighbor exchange couplings enables the 2 K peak to disappear. 
It should be stressed that the 30\% bond-randomness, which was estimated by Schnack {\it et al.} \cite{Schnack2} is needed to make an impact on the magnetization process, 
but the 10\% bond-randomness is enough to change the low-temperature specific heat curve, which originates from that fact that the singlet excitation gap, $\Delta_{\mathrm{s}}=0.048J$, of the Heisenberg model Eq.~(\ref{Heisenberg}) is a fraction  of the triplet excitation gap $\Delta_{\mathrm{t}}=0.218J$.
However, we have also found that the magnetic field dependence in specific heat curves still exists even though 10\% bond-randomness is incorporated.
In order to understand the experimental specific heat, it has been necessary to admit the existence of a bond-randomness of about 50\%.

Kihara {\it et al.} measured specific heat of another spherical kagom\'{e} system \{Mo$_{72}$V$_{30}$\}, together with that of \{W$_{72}$V$_{30}$\} \cite{Kihara}.
The thirty V$^{4+}$ with spin-1/2 in \{Mo$_{72}$V$_{30}$\} form an icosidodecahedron, as in \{W$_{72}$V$_{30}$\},
while the Heisenberg model for \{Mo$_{72}$V$_{30}$\} is considered to contain some distortions \cite{Kunisada}.
Kihara {\it et al.} reported that the specific heat of \{Mo$_{72}$V$_{30}$\} depends on the strength of magnetic field and can be reproduced quantitatively by the Heisenberg model with distortions,
which suggests that the bond-randomness in \{Mo$_{72}$V$_{30}$\} is rather small.
This fact may indicate the magnitude of bond-randomness depends on the sample preparation conditions seriously.
Through this study, it was found that the low temperature specific heat is a good probe for the bond randomness.
In the future, it is desired to control bond-randomness in sample preparation.

\section*{Acknowledgment}

We thank Dr. T. Kihara and Professor H. Nojiri for providing their unpublished data and fruitful discussions. The authors would like to thank the Supercomputer Center, Institute for Solid State Physics, University of Tokyo for the use of their facilities.

\let\doi\relax


\end{document}